# Generalizing the Mean and Variance to Categorical Data Using Metrics

Roger Bilisoly[1]

[1] Central Connecticut State University, 1615 Stanley Street, New Britain, CT 06050

**Abstract**

Researchers have developed ways to generalize the mean and variance to situations in which a data metric is available. We apply the tools developed in Pennec (2006) to categorical data, and show the generality of this approach by considering two quite different applications. First, spelling variability in Middle English is quantified. Second, variability of a finite group (in the sense of group theory) is defined and applied to an example.

## 1. Introduction

The usual sample mean and variance are appropriate for real-valued numerical data. However, data can also lie on Riemannian manifolds. For example, angular data such as longitudes of cities are naturally plotted on a circle, which is a simple manifold. Many other examples are considered in Fisher (1996) and Mardia and Jupp (2000). The key here is that angles (in radians) are not two dimensional points that lie approximately on a circle but are values that represent arc length from a labeled point and are inherently circular. Generalizations of standard statistical ideas such as mean and variance have been developed for manifolds in Pennec (1999) and Pennec (2006). It turns out that this approach can be extended to categorical data sets that have a metric associated with them. We show how this is done and illustrate it with two concrete examples: measuring spelling variability in Middle English and defining the variance of a finite group (in the sense of group theory).

## 2. Method

The usual mean can be defined in terms of minimizing the sum of squares in Equation (2.1) with respect to $c$. The minimizing value is the mean, call this $\mu$, and $f(\mu)/n$ is the population variance, as shown in (2.2).

$$f(c) = \frac{1}{n}\sum_{i=1}^{n}|x_i - c|^2 \qquad (2.1)$$

$$\text{variance} = \min_c f(c)/n \equiv f(\mu)/n \qquad (2.2)$$

JSM 2013 - Section on Statistical Learning and Data Mining

Section 4.1 of Pennec (2006) extends this idea to data on a manifold by replacing the absolute value sign, which is the distance function on the real numbers, with a geodesic distance on a manifold, denoted by the function *d* in Equation (2.3).

$$f(c) = \sum_{i=1}^{n} d(x_i, c)^2 \qquad (2.3)$$

$$\text{variability} = \min_c f(c). \qquad (2.4)$$

The generalization of the mean is the minimizing value(s) of *f*, and the generalized variance is the minimum value of *f* in Equation (2.4). Note that there need not be a unique solution, $c_{min}$. Moreover, there might be more than one geodesic through two points, so *d(x,c)* means the minimum arc length between *x* and *c* over all geodesics that connect these. Finally, for this definition to work, we must have a connected manifold with neither boundary nor singular points: details are in Pennec (2006).

Although we don't use the following ideas, Pennec (2006) goes on to define probability distributions and expectations on a manifold, which allows him to define covariance matrices, to generalize the normal distribution to a manifold by using maximum entropy, and to formulate a generalized $\chi^2$ law. All these are defined intrinsically as opposed to using definitions that embed the manifold in a higher dimensional space. For example, one can think of a circle as a one-dimensional manifold (an intrinsic point of view) or as a graph embedded in $R^2$. To make this distinction clear, an example of circular data variability is given below. This is important because intrinsic geometry in general relativity revolutionized physics. Since many statistical procedures can be viewed geometrically (for example, this is systematically done in Saville and Wood (1997)), and because the intrinsic methods of differential geometry have already been applied to statistics (for instance, see Murray and Rice (1993)), this could be a fruitful point of view for statisticians.

**2.1 Extending Pennec's Theory to Discrete Data**
There is nothing special about numerical data in Equations (2.3) and (2.4). The same formulas can be used if *x* and *c* are categorical data as long a metric function, *d*, is available. Moreover, squaring *d* corresponds to $L^2$-norm for data vectors, but other distance functions could be considered. For example, Equation (2.5) corresponds to the $L^1$-norm, and minimizing this *f* results in the median.

$$f(c) = \frac{1}{n}\sum_{i=1}^{n} |x_i - c| \qquad (2.5)$$

Because the median is more robust to outliers than the mean, we compare the results from Equations (2.3) and (2.4) to that of (2.6) and (2.7) in the examples below. In addition, although we do not use this, it is clear that a further generalization is possible by replacing the exponent 2 in Equations (2.1) and (2.3) with *p*, which is the $L^p$-norm of the data when considered as a vector.

$$f(c) = \sum_{i=1}^{n} d(x_i, c) \qquad (2.6)$$

$$\text{variability} = \min_c f(c) \qquad (2.7)$$

To illustrate the above theory, we consider three data examples. First, we consider a toy problem using a small data set to compare the methods discussed above. Second, we compute spelling variability using the Levenshtein edit distance, which is one of many text metrics. Finally, we compute finite group variability using the word metric from group theory.

## 2.2 Manifold Example with Circular Data

To illustrate Pennec (2006)'s generalization of mean and variance, we consider a data set consisting of angles, which can be viewed as values from a circle and is called *circular data* in Fisher (1996) or *directional data* in Mardia and Jupp (2000). Once this concrete example is understood, the generalizations to categorical data are straightforward.

Consider the angles {-2.12, -1.08, 0.016, 0.99, 2.08, 3.14}, which were generated by adding uniform noise to {-2π/3, -π/3, 0, π/3, 2π/3, π}. The traditional method of finding a mean direction is adding the unit vectors $e^{i\theta}$ for the values in the data set to produce a resultant vector. The solution is the angle this makes with the positive x-axis (unless the sum is 0, which gives no solution). In this case, the resultant is (0.0104, -0.00815), which gives an answer of -0.66 radians. Note that just averaging these angles as numbers gives the much different answer of 0.504 radians.

For a circle, the distance between two angles is given by Equation (2.8).

$$d(\theta_1, \theta_2) = \min(|\theta_1 - \theta_2|, 2\pi - |\theta_1 - \theta_2|) \qquad (2.8)$$

The minimization is needed because two angles create two arcs on the circle, and the shortest of these is the distance between the angles. For example, if $\theta_1$ = 1.5 and $\theta_2$ = -1.5, then the difference is 3, but the distance between them is π – 3 = 0.14.

We can minimize Equation (2.3) numerically to get a mean direction of -1.59. However, it is informative to look at a plot of (2.3), which is shown in Figure 1. Note that the data values correspond to the spikes, and all the local minimums are close to halfway between the data values. The y-coordinate of the global minimum is 19.0, which is the variability measure, although in practice one might divide this by the size of the data set. Finally, using Equation (2.6) produces an interval from -1.08 to -2.12. Since this is a generalization of the median, it is not surprising that a data set of even size produces a non-unique answer, which is not the case for an odd-sized sample.

Why are the answers above so different? First, treating angles as numbers is incorrect. Second, the six data points are almost {-2π/3, -π/3, 0, π/3, 2π/3, π}, which causes both the traditional resultant vector method and Equation (2.8) to be sensitive to small perturbations. This makes it is easy to find examples where these methods do not agree.

## 3. Applications to Discrete Data

As discussed in Section 2.1, the optimization approach to variability can be extended to any type of discrete data that has a metric. Metrics have been well studied in mathematics, statistics, and applications, so this approach has many uses, two of which are given below.

## 3.1 Spelling Variability Using Levenshtein Edit Distance

The history of English is split into three periods. First is Old English, which runs from roughly 450 (all dates are in the Common Era) through 1100, just after the Norman

Conquest. The Anglo-Saxon Chronicle states that the earliest settlers were the Angles, Saxons, and Jutes, which suggests that there were dialects in Britain from the start.

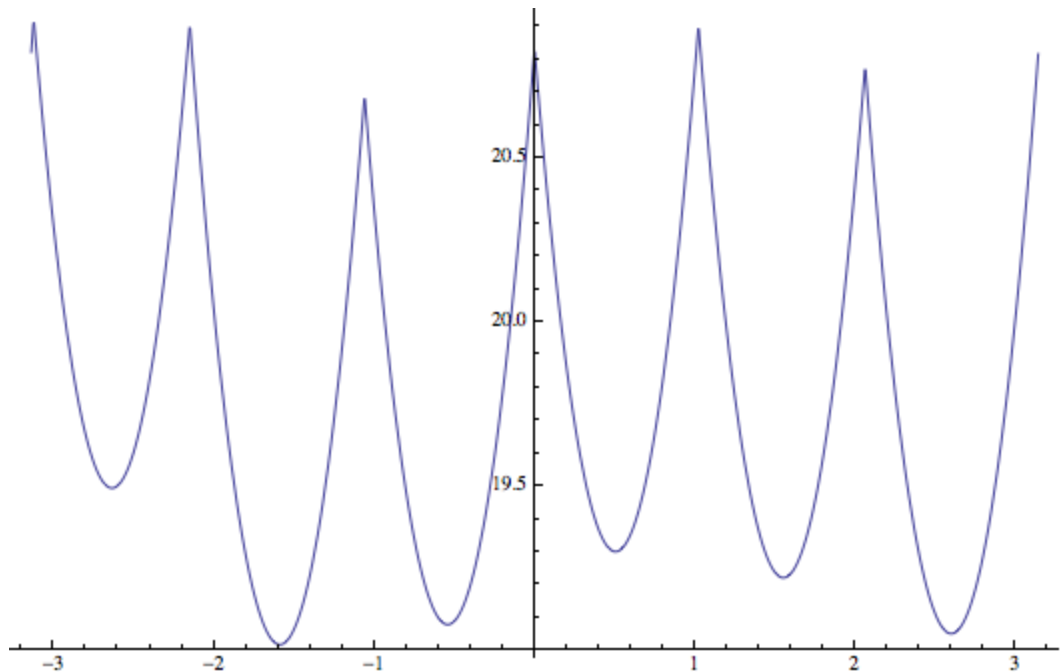

**Figure 1:** Plot of Equation (2.3) for the angle data in Section 2.2. The x-coordinate of the global minimum of this function gives the mean direction, while the y-coordinate is a measure of variability.

Middle English was used from about 1100 through 1500, ending just after the introduction of printing in England by William Caxton. It has several dialects, and it changed over time, both of which cause spelling variability. Figure 2 shows the first four lines of the General Prologue of The Canterbury Tales for four different manuscripts, and no two of these are identical. However, at that time, there were no dictionaries or other references that prescribed a standard orthography, so even within one manuscript spelling variations are common.

| | |
|---|---|
| Whan that Aprille . with his schoures swote. <br> The drought of Marche . hath perced to þe rote. <br> And bathed euery veyne . in suche licoure. <br> Of whiche vertue . engendrid ys the floure. <br> [Canterbury Tales, Cambridge MS] | WHan that Aprille with hise shoures soote <br> The droghte of March / hath perced to the roote <br> And bathed euery veyne / in swich licour <br> Of which vertu / engendred is the flour <br> [Canterbury Tales, Ellesmere MS] |
| WHan that Apprille / with his shouris soote <br> the drought of Marche / hath pershid to the roote <br> and bathed euery veyne in swich licoure <br> of which vertue / engendrid is the floure <br> [Canterbury Tales, Corpus MS] | WHan that Auerylle with his shoures soote <br> The droghte of March / hath perced to the roote <br> And bathed euery veyne in swich lycour <br> Of which vertu engendred is the flour <br> [Canterbury Tales, Hengwrt MS] |

**Figure 2:** Four different manuscripts of the General Prologue of *The Canterbury Tales* by Geoffrey Chaucer. No two of these have exactly the same spellings.

Third, Modern English starts around 1500, and by 1700 it is much like Present Day English. It is in this time period that modern dictionaries and grammars are developed and the idea of editorial standards take over.

Of these three periods, Middle English is the most variable, and it is well known that spelling variability decreases over time as book publishing and reference works become widely used. The question we address is how this variability can be explicitly quantified. As discussed in Section 2, we use a string metric to do this.

### 3.1.1 *Defining Levenshtein edit distance*

Levenshtein edit distance (we drop his name below) is defined to be the minimum cost of changing one string into another where letter copying has cost 0; adding, deleting, or substituting a letter all have cost 1. This can be computed by dynamic programming, which requires roughly $m*n$ steps, where $m$ and $n$ are the lengths of the two strings. Additional information is given in Chapter 6 of Russell (2011). For DNA matching in bioinformatics, this is computationally expensive because the strings can be long, but for English words, this is quick to compute.

The intuition behind this algorithm is to align two different strings as closely as possible, after which the letters that do not match are added, deleted, or substituted as needed. It turns out that it is enough to focus on initial substrings, starting with the empty string. Figure 3 shows an example where the edit distance between "OLD" and "HALDE" (a Middle English form of the word "OLD") is found to be 3. The optimal path is shown in red, and moving one square to the right means adding a letter, moving one square diagonally means a letter substitution, and moving one square downwards would mean deleting a letter. Intuitively, the cost is 3 because (1) "LD" is in both words and copying has no cost; (2) "O" is changed to "HA" at a cost of 2, and (3) "E" is added at the end at a cost of 1.

|   | "  | H | A | L | D | E |
|---|----|---|---|---|---|---|
| " | **0** | 1 | 2 | 3 | 4 | 5 |
| O | 1 | **1** | **2** | 3 | 4 | 5 |
| L | 2 | 2 | 2 | **2** | 3 | 4 |
| D | 3 | 3 | 3 | 3 | **2** | **3** |

**Figure 3:** The cost of changing "OLD" to "HALDE" is 3 as shown by the path of red numbers. Hence the edit distance between these two strings is 3.

It turns out that edit distance is a distance function in terms of the mathematical definition of metric space. That is, the following are true. First, EditDistance[s1, s2] ≥ 0 because all the costs are non-negative. Second, EditDistance[s1, s2] = 0 exactly when s1 = s2 because the only zero cost operation is copying. Third, EditDistance[s1, s2] = EditDistance[s2, s1] because (1) copying and substitution are their own inverses and (2) adding and deleting are inverses of each other. That is, for any transformation from s1 to

s2, this can be reversed to change s2 to s1 at the same cost. Fourth, EditDistance[s1, s2] + EditDistance[s2, s3] ≥ EditDistance[s1, s3] because, by definition, edit distance is the minimum cost over all string transformation paths, which includes paths going through s2.

### 3.1.1 *Variability of the Middle English forms of the word "OLD"*

Using edit distance, we now compute the variability of the forms of the word "OLD" appearing in the *Linguistic Atlas of Late Mediaeval English* (LALME), McIntosh et al. (1986). These are {aeld, aelde, ald, alde, alld, aulde, awlde, eeld, eelde, eld, elde, hald, halde, held, helde, hold, holde, hoolde, old, olde, oold, oolde, ould, wold, woold}. Each row and column of the matrix in Figure 4 stands for one of these words, and each entry is the respective edit distance. To find the median word using Equation (2.6), we find the row with the smallest sum, which gives two solutions: "hold" (16$^{th}$ row) and "old" (19$^{th}$ row) because both sum to 48. To find the mean word using (2.3) requires finding the row with the smallest sum of squares, which is 108 and corresponds to "hold."

$$\begin{pmatrix}
0 & 1 & 1 & 2 & 1 & 2 & 2 & 1 & 2 & 1 & 2 & 2 & 3 & 1 & 2 & 2 & 3 & 4 & 2 & 3 & 2 & 3 & 2 & 2 & 3 \\
1 & 0 & 2 & 1 & 2 & 1 & 1 & 2 & 1 & 2 & 1 & 3 & 2 & 2 & 1 & 3 & 2 & 3 & 3 & 2 & 3 & 2 & 3 & 3 & 4 \\
1 & 2 & 0 & 1 & 1 & 2 & 2 & 3 & 1 & 2 & 1 & 2 & 2 & 3 & 2 & 3 & 4 & 1 & 2 & 2 & 3 & 2 & 2 & 3 \\
2 & 1 & 1 & 0 & 2 & 1 & 1 & 3 & 2 & 2 & 1 & 2 & 1 & 3 & 2 & 3 & 2 & 3 & 2 & 1 & 3 & 2 & 3 & 3 & 4 \\
1 & 2 & 1 & 2 & 0 & 2 & 2 & 2 & 3 & 2 & 3 & 2 & 3 & 2 & 3 & 2 & 3 & 4 & 2 & 3 & 2 & 3 & 2 & 2 & 3 \\
2 & 1 & 2 & 1 & 2 & 0 & 1 & 3 & 2 & 3 & 2 & 3 & 2 & 3 & 2 & 3 & 3 & 2 & 3 & 2 & 2 & 3 & 4 \\
2 & 1 & 2 & 1 & 2 & 1 & 0 & 3 & 2 & 3 & 2 & 3 & 2 & 3 & 2 & 3 & 3 & 2 & 3 & 2 & 3 & 3 & 4 \\
1 & 2 & 2 & 3 & 2 & 3 & 3 & 0 & 1 & 1 & 2 & 2 & 3 & 1 & 2 & 2 & 3 & 4 & 2 & 3 & 2 & 3 & 2 & 2 & 3 \\
2 & 1 & 3 & 2 & 3 & 2 & 2 & 1 & 0 & 2 & 1 & 3 & 2 & 2 & 1 & 3 & 2 & 3 & 3 & 2 & 3 & 2 & 3 & 3 & 4 \\
1 & 2 & 1 & 2 & 2 & 3 & 3 & 1 & 2 & 0 & 1 & 2 & 3 & 1 & 2 & 2 & 3 & 4 & 1 & 2 & 2 & 3 & 2 & 2 & 3 \\
2 & 1 & 2 & 1 & 3 & 2 & 2 & 2 & 1 & 1 & 0 & 3 & 2 & 2 & 1 & 3 & 2 & 3 & 2 & 1 & 3 & 2 & 3 & 3 & 4 \\
2 & 3 & 1 & 2 & 2 & 3 & 3 & 2 & 3 & 2 & 3 & 0 & 1 & 1 & 2 & 1 & 2 & 3 & 2 & 3 & 2 & 3 & 2 & 2 & 3 \\
3 & 2 & 2 & 1 & 3 & 2 & 2 & 3 & 2 & 3 & 2 & 1 & 0 & 2 & 1 & 2 & 1 & 2 & 3 & 2 & 3 & 2 & 3 & 3 & 4 \\
1 & 2 & 2 & 3 & 2 & 3 & 3 & 1 & 2 & 1 & 2 & 1 & 2 & 0 & 1 & 1 & 2 & 3 & 2 & 3 & 2 & 3 & 2 & 2 & 3 \\
2 & 1 & 3 & 2 & 3 & 2 & 2 & 2 & 1 & 2 & 1 & 2 & 1 & 1 & 0 & 2 & 1 & 2 & 3 & 2 & 3 & 2 & 3 & 3 & 4 \\
2 & 3 & 2 & 3 & 2 & 3 & 3 & 2 & 3 & 2 & 3 & 1 & 2 & 1 & 2 & 0 & 1 & 2 & 1 & 2 & 1 & 2 & 2 & 1 & 2 \\
3 & 2 & 3 & 2 & 3 & 2 & 2 & 3 & 2 & 3 & 2 & 2 & 1 & 2 & 1 & 1 & 0 & 1 & 2 & 1 & 2 & 1 & 3 & 2 & 3 \\
4 & 3 & 4 & 3 & 4 & 3 & 3 & 4 & 3 & 4 & 3 & 3 & 2 & 3 & 2 & 2 & 1 & 0 & 3 & 2 & 2 & 1 & 3 & 3 & 2 \\
2 & 3 & 1 & 2 & 2 & 3 & 3 & 2 & 3 & 1 & 2 & 2 & 3 & 2 & 3 & 1 & 2 & 3 & 0 & 1 & 1 & 2 & 1 & 1 & 2 \\
3 & 2 & 2 & 1 & 3 & 2 & 2 & 3 & 2 & 2 & 1 & 3 & 2 & 3 & 2 & 2 & 1 & 2 & 1 & 0 & 2 & 1 & 2 & 2 & 3 \\
2 & 3 & 2 & 3 & 2 & 3 & 3 & 2 & 3 & 2 & 3 & 2 & 3 & 2 & 3 & 1 & 2 & 2 & 1 & 2 & 0 & 1 & 1 & 1 & 1 \\
3 & 2 & 3 & 2 & 3 & 2 & 2 & 3 & 2 & 3 & 2 & 3 & 2 & 3 & 2 & 2 & 1 & 1 & 2 & 1 & 1 & 0 & 2 & 2 & 2 \\
2 & 3 & 2 & 3 & 2 & 2 & 3 & 2 & 3 & 2 & 3 & 2 & 3 & 2 & 3 & 3 & 1 & 2 & 1 & 2 & 0 & 2 & 2 \\
2 & 3 & 2 & 3 & 2 & 3 & 3 & 2 & 3 & 2 & 3 & 2 & 3 & 1 & 2 & 3 & 1 & 2 & 1 & 2 & 2 & 0 & 1 \\
3 & 4 & 3 & 4 & 3 & 4 & 4 & 3 & 4 & 3 & 4 & 3 & 4 & 3 & 4 & 2 & 3 & 2 & 2 & 3 & 1 & 2 & 2 & 1 & 0
\end{pmatrix}$$

**Figure 4:** The edit distances between each pair of word forms for "OLD" in the LALME.

The above method can now be used to compare variabilities. For example, according to Chaucer et al. (1987), Chaucer only uses four forms: {olde, old, oold, oolde}. The matrix of distances is a submatrix of Figure 4, and one finds that all the words are means as well as medians. The variability produced by Equation (2.6) is 4, and by (2.3) is 6, so Chaucer is less variable than LALME, which is expected since his texts are a subset of the latter's.

## 3.2 Variability of Finite Groups Using the Word Metric

A group is a set with a binary associative operation that satisfies the following: (1) the set is closed under the operation; (2) there is an identity element; and (3) every element has an inverse. For example, the set {0, 1, …, $n$-1} with the operation addition modulo $n$ is a group. And the set {1, 2, …, $p$-1} with the operation multiplication modulo $p$ is a group when $p$ is prime, which is required so that there are multiplicative inverses modulo $p$. For more on this definition, see any introductory text on abstract algebra such as Chapter 1 of Fraleigh (2003).

Groups can be defined with generators and a set of relationships, an approach called combinatorial group theory: see Miller (2004). For example, {0, 1, …, $n$-1} with addition modulo $n$ is generated by 1. That is, repeatedly adding 1 generates all these values. It is also generated by -1 and any other value relatively prime to $n$. Another example is {1, 2, …, $p$-1} with multiplication modulo a prime, $p$, which is generated by any primitive root, a concept from number theory. A last example is the dihedral group of size 2n, which has two generators that satisfy $a^n = 1$, $b^2 = 1$, and $(ab)^2 = 1$. In general, let $G$ be the group, and let $S$ be a set of generators, and assume that $S$ is closed under inverses, which is not required, but simplifies matters below.

There is a standard distance for a group presented as generators and a list of identities, which is called the *word metric*. This is given by Equation (3.1).

$$d(a,b) = \min_d b = s_1 s_2 \ldots s_d a, \qquad (3.1)$$

where the $s_i$ are in the generating set $S$. Note a similarity with edit distance: both use the minimum number of operations to change one object into another.

The easiest way to compute word distances is to create a Cayley graph using the generator set, $S$, and then use the fact that the usual distance between two vertices on this graph is the same as the word distance just defined. Figure 5 does this for the cyclic group of order 15, $C_{15}$, where S = {1, -1}, and Figure 6 does this for the direct product of cyclic groups, $C_3$ x $C_5$, where S = {{1, 0}, {-1, 0}, {0, 1}, {0, -1}}. Finally, notice that $C_{15}$ and $C_3$ x $C_5$ are isomorphic, but because we are using two different sets of generators, the graphs in Figures 5 and 6 are different.

For each matrix in Figure 7, the rows have the same numbers, just rotated. So by Equation (2.6), every vertex is a median, and the variability is the sum of any row, which is 56 for $C_{15}$ and 28 for $C_3$ x $C_5$. By (2.3), every vertex is a mean, and the variability is the sum of the squares of each row, which is 280 for $C_{15}$, and 64 for $C_3$ x $C_5$. Since each group has order 15 (the number of vertices), directly comparing these values is valid. Note that graph theory uses the concept of average distance, and this corresponds to (2.6) for this example because the average is a linear function of the sum.

Finally, the above method depends critically on the choice of generators, $S$. An extreme example is using $S = G$, then for all $a$ and $b$ distinct, $d(a,b) = 1$ because $s = ba^{-1}$ can be used in Equation (3.1). This makes the Cayley graph the complete graph, so any two groups of equal size would be equal in variability. In the above cases, natural sets of generators were used, so $C_3$ x $C_5$ is less variable than $C_{15}$.

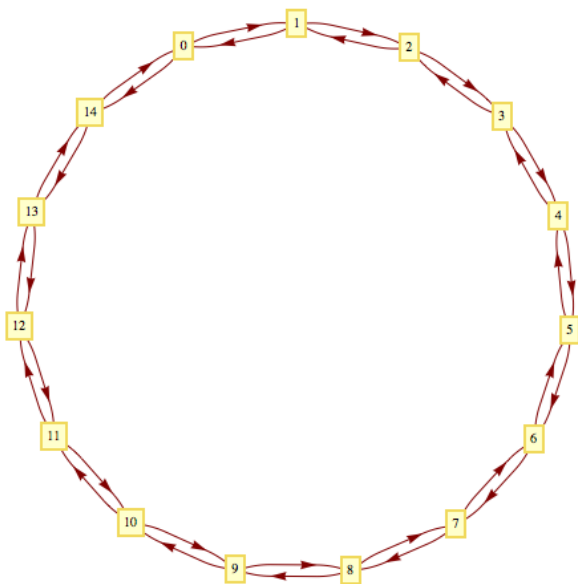

**Figure 5:** The Cayley graph of $C_{15}$ (using addition modulo 15) with the two generators $\{1, -1\}$.

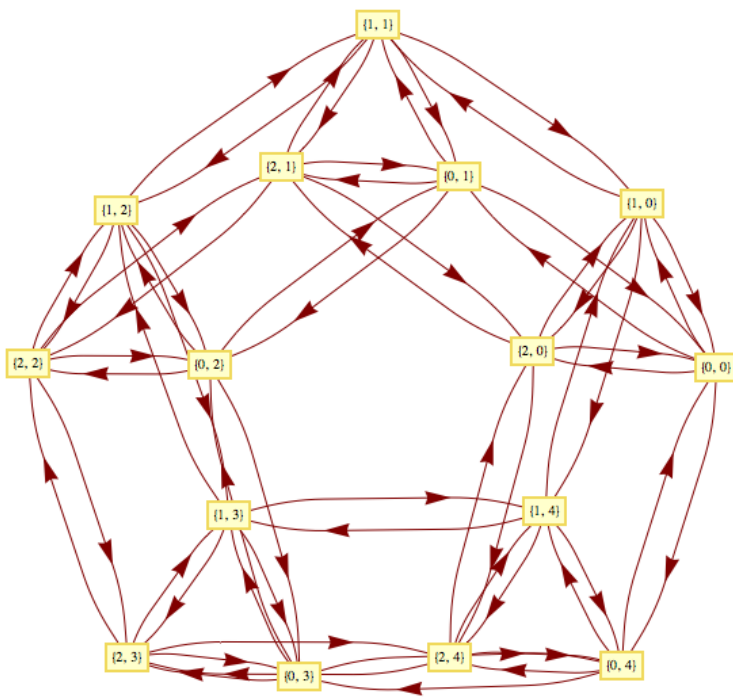

**Figure 6:** The Cayley graph of $C_3$ x $C_5$ (using addition modulo 3 and 5, respectively) with the four generators $\{\{1, 0\}, \{-1, 0\}, \{0, 1\}, \text{and } \{0, -1\}\}$.

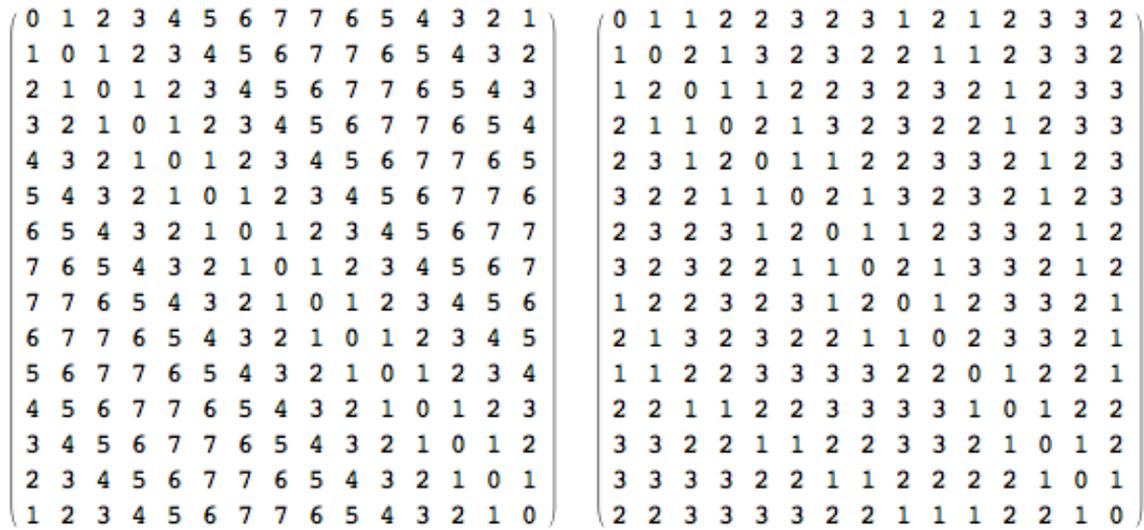

**Figure 7:** These are the distance matrices for $C_{15}$ and $C_3 \times C_5$, respectively. Note that the entries in the latter are mostly smaller than those in the former.

## 4. Discussion

The Equations (2.3) and (2.4) as well as (2.6) and (2.7) allow the definition of a typical value and variability for discrete data. Because much basic statistical theory is based on means and variances, the above generalizations can also be used to create hypothesis tests. For example, the F test, which uses the ratio of variances, is one way to decide whether or not $H_0$: $\sigma_1^2 = \sigma_2^2$ is true. This suggests using Equation (4.1) as a generalization.

$$\frac{\sum_{i=1}^{m} d(x_i, c_{min})^2}{\sum_{j=1}^{n} d(y_j, c_{min})^2} \tag{4.1}$$

The distribution of (4.1) is unknown in general, but this could be estimated by a permutation test. That is, take a random sample of size $m$ from $\{x_1, x_2, ..., x_m, y_1, y_2, ..., y_n\}$, use this sample as the $x$s, and the rest as the $y$s. Repeating this many times gives an empirical sampling distribution, which can be used to estimate the p-value.

Finally, there are many metrics that already exist, some of which already have known uses in statistics. For instance, Diaconis (1988) discusses many metrics of the symmetric group, $S_n$, and these are related to rank-based statistical techniques. For example, he shows that Spearman's rank correlation is equivalent to the $L^2$-norm on $S_n$. The generality of the above approach will make further applications to categorical data easy to find.

## Acknowledgements

Thanks to my colleagues Krishna Saha for many discussions on the statistical analysis of discrete data and Fred Latour for discussions on abstract algebra.